\documentclass[12pt]{article}

\usepackage{graphicx,epsf}

\def\beq{\begin{equation}}
\def\eeq{\end{equation}}
\def\mP{m_{\rm P}}
\def\d{\delta}

\def\in{{(i)}}
\newcommand{\gsim}{\ \raise.3ex\hbox{$>$\kern-.75em\lower1ex\hbox{$\sim$}} \ }
\newcommand{\lsim}{\ \raise.3ex\hbox{$<$\kern-.75em\lower1ex\hbox{$\sim$}} \ }
\newcommand{\bea}{\begin{eqnarray}}
\newcommand{\eea}{\end{eqnarray}}

\title{Mixed  inflaton and curvaton perturbations}

\author{David Langlois and  Filippo Vernizzi\\
{\small {\it GReCO, Institut d'Astrophysique de Paris, CNRS,}}\\
{\small {\it 98bis Boulevard Arago, 75014 Paris, France}}}

\date{\today}

\begin{document}

\maketitle

\abstract{
A recent variant of the inflationary paradigm is that the ``primordial''
curvature perturbations come from quantum fluctuations of a scalar 
field, subdominant and effectively massless during inflation, called 
the ``curvaton'', instead of the fluctuations of the inflaton field. 
We consider the situation  where the primordial
curvature perturbations  generated
by the quantum fluctuations of  an 
inflaton and of a curvaton field are of the same order of magnitude.
We compute 
the total curvature perturbation and
its spectrum in this case and we discuss the observational
consequences.
}

\section{Introduction}
The first observations of the WMAP satellite \cite{WMAP1} have
confirmed the basic predictions of the inflation scenarios
\cite{guth} (see also \cite{liddle_lyth}), namely the existence of
super-Hubble primordial fluctuations, mostly adiabatic and
characterized by a quasi-scale-invariant spectrum.
Because direct information on the primordial universe is scarce, it is
essential, when trying to interpret the cosmological data, to determine
to which extent a precise measurement of the
primordial spectrum gives us information on the inflationary scenario
itself.

In this perspective, whereas in the standard inflationary
paradigm, the primordial perturbations are generated by 
quantum fluctuations of the {\it inflaton}, it has been realized
recently that the observed primordial fluctuations could be
accounted for by quantum fluctuations of a scalar field other
than the inflaton, dubbed the {\it curvaton} field
\cite{lw02,es02,mt01} (and see also \cite{luw} and \cite{mt02}).
The curvaton is a scalar field $\sigma$ 
whose effective mass must be much smaller 
than the Hubble parameter during inflation so that it acquires 
quantum fluctuations with a quasi scale-invariant spectrum. 
After inflation, the primordial curvature  perturbations
are generated by the conversion of
the isocurvature perturbations due to the fluctuations of the   
subdominant curvaton field, 
into adiabatic perturbations.

In the curvaton scenario, inflation is still necessary since this is the 
source of the curvaton quantum fluctuations but the separation of the 
r\^oles played by the inflaton that drives the evolution of the universe and
by the curvaton which generates the perturbations, liberates the inflaton 
from some observational constraints. 
This alternative possibility is mixed news for the inflationary
scenarios. It is good news in the sense that it allows to construct
more easily explicit inflationary models with a firmer footing
from a particle physicist's point of view \cite{Lythmodels}. But
it is also bad news in the sense that it shows a new limitation to
extract information on the nature of the inflaton field from the
measurement of the primordial fluctuations.

 In the curvaton
scenario, the curvature  fluctuations generated by the inflaton are usually
assumed to be negligible, and the primordial fluctuations are thus essentially 
due to the curvaton. 
However, as was pointed out in \cite{dlnr03}, one can envisage  cases 
where   the fluctuations generated by  both the inflaton   and
a curvaton-like field are relevant. This is the situation that we explore
in the present work.

It can occur, for instance,   when  $\sigma_*$ is  of the order of
the Planck mass $\mP$ during inflation, still assuming that the curvaton energy density 
is much smaller than that of the inflaton (otherwise we would be in the 
context of double inflation \cite{ps92}), and  implies that the 
curvaton starts  to oscillate 
when its  energy density  already contributes significantly 
to the total energy density. This is in contrast with the usual curvaton
scenario where the curvaton starts oscillating long before it
dominates and  can thus be treated as a pressureless fluid during
the first radiation dominated phase. 

Curvaton and inflaton generated perturbations can also be of the same
order of magnitude for  small $\sigma_*$, i.e.\ $\sigma_*\ll \mP$, but 
with a slow-roll inflation parameter $\epsilon$ sufficiently small during
inflation to produce big inflaton perturbations.

This paper is organized as follows. In the next section we write
the background evolution equations and we derive their initial
conditions explaining the context of our present analysis. In
Sec.~III we discuss the perturbation equations and their initial
conditions. We can analytically derive the curvature perturbation
after curvaton decay, i.e.\ at the onset of the standard big bang
cosmology, in two limits. In the pure curvaton limit, for
$\sigma_* \ll \mP$, we recover the curvature perturbation 
already derived in previous works.
 In the limit where the curvaton
acts as a secondary inflaton, for $\sigma_* \gsim \mP$, we recover
the perturbation of the double inflation model. In the
intermediate case we resort to a  numerical analysis to derive the
mixed inflaton and curvaton perturbation, which is the main result
of our work. In Sec.~IV we derive the spectrum of the curvature
perturbation, its spectral index, and discuss the observational
consequences of the model.
Finally we conclude in the last section.

\section{Homogeneous equations of motion}
In the curvaton scenario, the inflationary phase 
 is followed by a {\it first}
radiation dominated era, where the dominant species are the decay products
of the inflaton. The curvaton is a scalar field,  $\sigma$, 
that 
is  assumed to be subdominant during the whole
inflationary epoch and at the beginning of the radiation dominated
post-inflationary era. 
Its effective mass is assumed to be much smaller than the Hubble 
parameter during inflation. Here, for simplicity, we take the 
quadratic potential 
\beq
V(\sigma)={1\over 2}m^2\sigma^2.
\eeq

In the  post-inflationary  era, two components coexist: the  radiation fluid and  
the curvaton.  
The evolution of the homogeneous background is thus governed by
the Friedmann equation
\beq
\label{friedmann}
H^2={8\pi G\over 3}\left(\rho_r+{1\over 2}\dot\sigma^2+{1\over 2}m^2 \sigma^2
\right),
\eeq
the continuity equation for the radiation energy density $\rho_r$,
\beq
\dot\rho_r+4H\rho_r=0,
\eeq
and
the equation of motion for the curvaton,
\beq
\ddot\sigma+3H\dot\sigma+m^2\sigma=0. \label{curvaton_background_eq}
\eeq

When $H\gg m$ the cosmological friction is so strong that the curvaton
is essentially frozen. This is the case  during
inflation since the curvaton must be effectively massless to acquire
the usual scale-invariant quantum fluctuations.
After inflation, $H$ keeps decreasing  until
it reaches the value $H\sim m$, at which point the curvaton  starts
oscillating.

In the standard curvaton scenario, $\sigma$ is supposed to start
oscillating while it is still strongly subdominant with respect to
radiation. Using (\ref{friedmann}), this implies 
\beq 
H^2\sim m^2
\gg {m^2\over \mP^2}\sigma_* ^2, 
\eeq 
where we have introduced the reduced Planck mass $\mP\equiv (8\pi G)^{-1/2}$.
This relation  means that the initial
amplitude of the curvaton must satisfy $\sigma_* \ll \mP$.
Conversely,  assuming an initial value $\sigma_* \sim \mP$ during
inflation implies that the epochs of oscillation and domination
roughly coincide for the curvaton. 

During the radiation dominated era, when $\sigma$ is subdominant,
$H=(2t)^{-1}$, and the curvaton equation of motion 
(\ref{curvaton_background_eq}) reads 
\beq \ddot
\sigma+{3\over 2 t}\dot\sigma+m^2\sigma=0. 
\label{curv_eom}
\eeq 
One recognizes a
Bessel equation, whose non-decaying solution is given by 
\beq
\label{sigma_RD1} \sigma_\in=\sigma_*\,   A \, { J_{1/4}(mt)\over
(mt)^{1/4}}, \qquad A\equiv {\pi\over 2^{1/4} \Gamma(3/4)}, 
\eeq
where $J_\nu$ stands for a Bessel function of rank $\nu$. In the
limit $mt\ll 1$, one gets $\sigma_\in\simeq \sigma_*$. The
expression $\sigma_{(i)}$ given above embodies the initial
conditions for the curvaton field.

Note  that the solution (\ref{sigma_RD1}) differs from the slow-roll 
solution that would apply during inflation when the Hubble parameter 
is slowly varying. Here, the acceleration is not small with respect to the
other terms in the equation of motion (\ref{curv_eom}). 
In particular,  in the limit
$m t \ll 1$, one finds \beq H {\dot\sigma_\in}\simeq -{1\over
5}m^2\sigma_\in. \eeq

\section{Analysis of the perturbations}

\subsection{Equations of motion}

The equations governing the evolution of the scalar-type 
perturbations are well-known and can be found in  review  papers such as
 \cite{ks} or \cite{mfb}. 
We adopt here 
 the {\it longitudinal gauge} in which  the perturbed metric reads 
\beq
ds^2=-(1+2\Phi)dt^2+a^2(t)(1-2\Psi)\d_{ij}dx^idx^j. 
\eeq
The perturbation of the curvaton field is denoted $\d\sigma$ and
the perturbations of the radiation fluid are described by the 
perturbation of the energy density, $\d\rho_r$, or the density contrast
$\d_r=\d\rho_r/\rho_r$, and the peculiar velocity $v_r$.

Taking into account the relation $\Psi=\Phi$, imposed by the
Einstein equations in absence of anisotropic stress,
the perturbed equations of motion  are 
the energy and momentum constraints  in Einstein's equations,
respectively 
\beq \label{energy_constraint} -3H(H\Phi + \dot \Phi)
- \frac{k^2}{a^2}\Phi = 4 \pi G \left( \dot \sigma \dot{\delta
\sigma} + m^2\sigma \delta \sigma- \dot \sigma^2 \Phi + \rho_r
\delta_r \right) \eeq 
and 
\beq \dot\Phi+H\Phi=4\pi G\left(
\dot\sigma \d \sigma-{4\over 3}\rho_r v_r \right) , 
\eeq 
the
continuity and Euler equations for the radiation fluid,
respectively 
\beq \dot\delta_r-{4\over 3}{k^2\over
a^2}v_r-4\dot\Phi=0 \eeq 
and 
\beq \dot v_r-H v_r+\Phi+{1\over
4}\delta_r=0, \eeq
 and finally the Klein-Gordon equation for the curvaton,
\beq
\ddot{\d\sigma}+3H\dot{\d\sigma}+\left({k^2\over a^2}+m^2\right)
\d\sigma
=4\dot\sigma \dot\Phi-2m^2\sigma \Phi.
\eeq

Since the cosmological perturbations of observational interest today 
were on  super-Hubble scales  in the early universe, we will be interested
only in the long wavelength limit $k/(aH) \to 0$ of the above equations, 
which reduce to the 
following system:
\beq
\label{energy_constraint2}
-3H(H\Phi + \dot \Phi) = 4 \pi G \left(
\dot \sigma \dot{\delta \sigma} + m^2\sigma \delta \sigma- \dot \sigma^2
\Phi + \rho_r \delta_r \right),
\eeq
\beq \dot\delta_r-4\dot\Phi=0, \eeq
 and
\beq
\ddot{\d\sigma}+3H\dot{\d\sigma}+m^2
\d\sigma
=4\dot\sigma \dot\Phi-2m^2\sigma \Phi, \label{field_perturbation_eq}
\eeq
where we have gotten rid of the perturbed velocity $v_r$.

\subsection{Initial conditions}

Our initial conditions are  defined early in  the first radiation
dominated era
when the curvaton is at rest and subdominant. 
In this limit  the Bardeen potential $\Phi$ is constant. 
In  the standard
curvaton case,  it is further assumed to be zero, but here 
 it  cannot be neglected since it represents the contribution of the 
inflaton to the ``primordial'' curvature perturbation. 
Our initial conditions are thus given by
\beq
\label{init_conds}
\Phi=\Phi_*, \quad \dot\Phi=0, \quad \d\sigma=\d\sigma_*, \quad
\dot \d\sigma=0,
\eeq
where $\Phi_*$ stands for the curvature perturbation generated during 
inflation. 
Using the energy constraint (\ref{energy_constraint2}),
these initial conditions imply 
\beq
\d_r=-6 {m_P^2H^2\over \rho_r}\Phi_*-{m^2\sigma\over \rho_r}\d\sigma_*.
\eeq
Moreover, 
since the curvaton is initially subdominant,  the second term on the
right hand side can be dropped
and the initial condition for $\d_r$ reduces to
\beq
\label{dr_i}
\d_r^\in=-2\Phi_*.
\eeq

 The evolution of $\d\sigma$ is 
governed by the equation of motion  Eq.~(\ref{field_perturbation_eq}),
which can be approximated in our limit  by
\beq
\ddot{\d\sigma}+{3\over 2t}\dot{\d\sigma}+m^2
\d\sigma
=-2m^2\Phi_*\sigma.
\eeq
One recognizes the same Bessel equation as for the background but now
with a source term on the right hand side, which is proportional to
the background solution for $\sigma$ given in Eq.~(\ref{sigma_RD1}).
The solution for the perturbation is then found to be given by
\beq
\delta \sigma_\in = \frac{A}{ (mt)^{1/4}} \left\{
\left(\d\sigma_*-{1\over 2}\Phi_*\sigma_*\right) J_{1/4} (mt) +
\Phi_*\sigma_* mt
 J_{-3/4}(mt) \right\}.
\label{delta_sigma} \eeq 
Note also that using
Eq.~(\ref{sigma_RD1}) we can rewrite Eq.~(\ref{delta_sigma}) as
\beq 
\delta \sigma_\in = \frac{\delta \sigma_*}{\sigma_*}
\sigma_\in + t\dot \sigma_\in \Phi_*\  .
\label{delta_sigma_RD1}
\eeq
This yields $\delta \sigma_\in=\delta \sigma_* +{\cal O}(m^2 t^2)$, which 
justifies the initial conditions for the curvaton perturbation given
in (\ref{init_conds}).

As the energy density perturbation for the curvaton is given
explicitly by 
\beq 
\d\rho_\sigma=m^2 \sigma \delta \sigma + \dot
\sigma \dot{\delta \sigma} - \dot \sigma^2 \Phi, 
\eeq 
we can compute, using 
 (\ref{delta_sigma_RD1}),  the initial density contrast 
 \beq
\label{dsigma_in} \delta_\sigma^\in = 2 \frac{\delta
\sigma_*}{\sigma_*} - \frac{3}{2} \left. \frac{\dot
\sigma^2}{\rho_\sigma} \right|_\in \Phi_*\, .  
\eeq

\subsection{Evolution of the perturbations}

The above analytical expressions are  valid only during the early
radiation dominated era when the curvaton contribution to the
background energy density is negligible. This 
contribution however keeps increasing with time 
until it eventually becomes significant, and a more general analysis is 
required.  

The purpose of this section is to compute the curvature
perturbation after the curvaton decay, i.e.\ to establish the
initial conditions
 at the start of the {\it second radiation dominated era}, 
which must be identified  with
 the usual  radiation era of the standard cosmological scenario.
Since we  assume the decay to be instantaneous (it has been
shown recently that this is a very good approximation \cite{wmu}),
it is enough for our purpose to determine the curvature
perturbation when the curvaton is both dominating and oscillating, 
which will be denoted $\Phi_f$.
The corresponding curvature perturbation in the subsequent
radiation  phase is then simply obtained by applying the
usual transfer coefficient between a matter dominated phase and a
radiation dominated phase.

In order to determine the {\it final} curvature perturbation in
our scenario, i.e.\ the {\it primordial} perturbation for the
standard cosmological model, we have solved numerically the  system of
equations that consists of (\ref{friedmann})-(\ref{curvaton_background_eq}) 
for the background 
and of (\ref{energy_constraint2})-(\ref{field_perturbation_eq}) 
for the perturbations.
 For the homogeneous system, the only parameter that can
be varied is the initial amplitude $\sigma_*$ of the curvaton. As
for the perturbations, the initial conditions depend on two
parameters, the curvature perturbation $\Phi_*$ and the curvaton
perturbation $\delta\sigma_*$. Since this is a linear problem, the
final result can be written, quite generally as 
\beq
\label{transfert} \Phi_f=\alpha\Phi_*+\beta \d\sigma_*, 
\eeq 
where
the two coefficients $\alpha$ and $\beta$, to be determined,
depend only on the homogeneous quantities. This means that one can
analyze independently the cases $\Phi_*=0$ and $\d\sigma_*=0$, the
general result being given by the sum of these particular cases.

Let us start with the case $\d\sigma_*=0$. We will  now show that this
case corresponds to a purely adiabatic initial condition.
Let us first recall that the so-called isocurvature perturbation
between two matter components (here radiation and curvaton)
 can be defined as (see e.g. \cite{wmu})
\beq 
S \equiv 3(\zeta_r - \zeta_\sigma), 
\eeq 
where for a given
matter species $X$, $\zeta_X$ is the curvature perturbation on 
uniform energy density hypersurfaces, 
namely
\beq \label{zeta}
\zeta_X\equiv-\Phi-H{\d\rho_X\over \dot\rho_X}= -\Phi
+\frac{\delta_X}{3(1+w_X)}, 
\eeq 
with $w_X=P_X/\rho_X$. For the
curvaton, the pressure is $P_\sigma=(\dot\sigma^2-m^2\sigma^2)/2$
and the energy density is
$\rho_\sigma=(\dot\sigma^2+m^2\sigma^2)/2$ so that the entropy
perturbation is given by 
\beq 
S= \frac{\delta_r}{1+w_r} -
\frac{\delta_\sigma}{1+w_\sigma} = \frac{3}{4} \delta_r -
\frac{\rho_\sigma}{\dot \sigma^2 } \delta_\sigma. \label{entropy}
\eeq 
Substituting the initial conditions (\ref{dr_i}) and (\ref{dsigma_in}), one 
gets for the 
{\it initial} entropy perturbation 
 \beq 
\label{S}
S^\in=- 2\frac{\rho_\sigma}{\dot
\sigma^2 }\frac{\delta \sigma_*}{\sigma_*}. 
\eeq 
 Therefore, $\delta
\sigma_*=0$ corresponds to $S^\in=0$, i.e.\ to a purely
``adiabatic'' initial perturbation.

The perturbation then remains adiabatic throughout the evolution from
the radiation dominated era to the oscillating curvaton  dominated era.
Note that, in contrast with the case of two uncoupled
adiabatic perfect fluids where the individual $\zeta_X$ are separately
conserved, the entropy perturbation does not in general remain
constant  
on super-Hubble scales because  a non zero  intrinsic entropy of the
scalar field implies $\dot \zeta_\sigma \neq 0$.
However, if the initial entropy vanishes it will remain
so subsequently. Since we have a pure adiabatic perturbation,
the total $\zeta$ is conserved on super-Hubble  scales.
Consequently the curvature perturbation is simply governed by
the global equation of state of the matter content, or in other words
the total $\zeta$ is conserved in time. During a cosmological
 era with $w=$ const, the Bardeen potential $\Phi$ is constant and
is related to $\zeta$ via the relation (see e.g. \cite{mfb}) 
\beq
\label{zeta_Phi} \zeta=-{5+3w\over 3(1+w)}\Phi. 
\eeq 
Using this
relation both in the initial radiation dominated era with $w=1/3$
and in the oscillating curvaton dominated era with $w=0$, the
conservation of $\zeta$ implies that the coefficient
 $\alpha$ in Eq.~(\ref{transfert}) is given by 
\beq
\alpha={9\over 10}.
\eeq
We have  checked numerically that this result indeed
holds for all types of initial conditions 
(not only $\delta
\sigma_*=0$).

To determine the other coefficient, $\beta$, it is now sufficient
to consider the initial configurations with  $\Phi_*=0$. We
distinguish below several cases: first the limiting case where
$\sigma$ behaves as a standard curvaton, i.e.\ oscillates long
before domination (see upper part of Fig.~1); then the other
limiting case where $\sigma$ starts oscillating only after it
completely dominates and thus generates a second inflationary
phase (see upper part of Fig.~2);  and finally the general
intermediate case, as illustrated in the upper part of Fig.~3.

\subsubsection{The standard curvaton limit}

This is the limit corresponding to $\sigma_* \ll \mP$. As discussed before,
this implies that  the curvaton oscillates long before its domination.
One thus recovers the standard curvaton scenario, in which case the curvaton
can be  identified  with a pressureless fluid during the post-inflation phase.

\begin{figure}
\begin{center}
\includegraphics[width=4.8in]{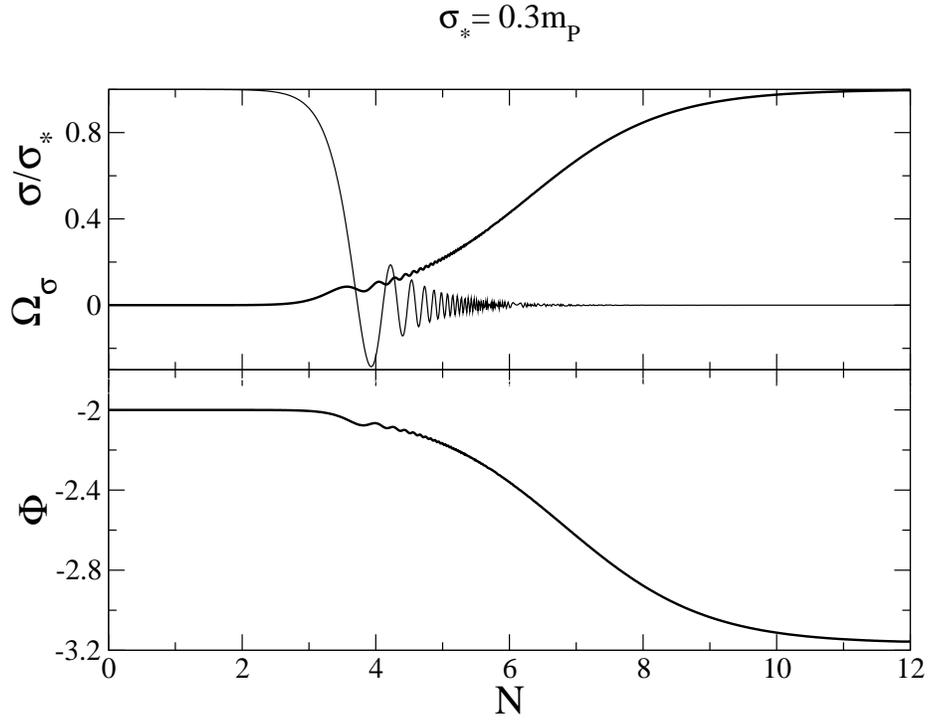}
\end{center}
\caption{The standard curvaton limit, for $\sigma_* = 0.3 \mP$.
Evolution of the curvaton, its energy density fraction $\Omega_\sigma\equiv 
\rho_\sigma/(\rho_\sigma+\rho_r)$ (upper
part), and of the curvature perturbation $\Phi$ (lower part). The
initial conditions for the perturbations
are $\delta \sigma_* = \mP$ and $\Phi_* =-2$.}
\end{figure}

Let us briefly recall the usual analysis in the standard curvaton
scenario \cite{lw02}. The curvaton is assimilated to a dust-like
fluid with density contrast $\d_m$. Since the two fluids are
decoupled, $\zeta_r$ and $\zeta_m$ are separately conserved,
whereas the total $\zeta$ is given by 
\beq
\zeta={4\rho_r\zeta_r+3\rho_m\zeta_m\over 4\rho_r+3\rho_m}. 
\eeq
Assuming no initial curvature perturbation, i.e.\ $\Phi_*=0$ and
thus $\zeta_r=0$, one finds that during the curvaton domination 
\beq
\label{zeta_d_m} \zeta=\zeta_m={1\over 3}\d_m^\in, 
\eeq 
where the
second equality comes from the definition of $\zeta$ given in
Eq.~(\ref{zeta}). Using the relation (\ref{zeta_Phi}) between
$\zeta$ and $\Phi$ during a $w=0$  era, one gets 
\beq
\Phi_f=-{3\over 5}\zeta= -{1\over 5}\d_m^\in. 
\eeq 
The final step  is to relate  the density contrast to the initial curvaton
fluctuation, by using  Eq.~(\ref{dsigma_in}). For
$\Phi_*=0$, this is simply $\d_\sigma^\in=2\d\sigma_*/\sigma_*$.
Therefore, in the standard curvaton limit, one finds for the
coefficient $\beta$ of Eq.~(\ref{transfert}) the value \beq
\beta=-{2\over 5\sigma_*}. \eeq

Although the expression for $\alpha$ has already been obtained by
the general argument given earlier, it is nevertheless instructive
to see how this value is recovered in the curvaton limit by
considering a non vanishing $\Phi_*$. When averaged over several
oscillations,  Eq (\ref{dsigma_in}) yields 
\beq \delta_m^\in
\equiv \delta_\sigma^\in = 2 \frac{\delta \sigma_*}{\sigma_*} -
\frac{3}{2} \Phi_*\, . \label{i_matter_density_constrast} \eeq 
During
the curvaton dominated era, one finds (note the difference with
Eq.~(\ref{zeta_d_m})) 
\beq \zeta=\zeta_m=-\Phi_* + {1\over
3}\d_m^\in. \eeq 
Substituting in $\Phi_f=-(3/5)\zeta$, derived
from (\ref{zeta_Phi}), one finally gets the expected result 
\beq
\label{transfert_curvaton} \Phi_f={9\over 10}\Phi_* -{2\over
5}{\d\sigma_*\over \sigma_*}. 
\eeq 
We have also checked this result by
evolving numerically the perturbation equations. As an
illustration, we show the evolution of  $\Phi$ in Fig.~1 for 
initial perturbations $\delta\sigma_*=\mP$ and $\Phi_*=-2$
\footnote{Within perturbation theory, the overall 
amplitude of the perturbations 
can be arbitrarily rescaled and we have chosen to set $\delta \sigma_*=\mP$.
Moreover, the initial condition $\Phi_*=-2$
corresponds to the curvature perturbation produced by a chaotic
quadratic model with $\phi_* =6 \mP$ and $\delta\phi_*=\mP$.}.

Note finally that the above  equation can be interpreted 
 as a particular case of
the more general equation
\beq
\label{transfert2}
\Phi_f={9\over 10}\Phi_* +{1\over 5}S,
\eeq
relating the final curvature perturbation, in a matter dominated era,
to the initial curvature perturbation and entropy perturbation defined
in a radiation dominated era. Eq.~(\ref{transfert2})  can be  easily derived
from $\Phi_f=-(3/5)\zeta_m$ by substituting $\zeta_m=\zeta_r- S/3$, which
follows from (\ref{entropy}), with
$\zeta_r=-\Phi_*+\d_r^\in/4=-(3/2)\Phi_*$.
The  expression (\ref{transfert2}) 
can be used only when the entropy perturbation
is conserved on large scales. This is the case for the curvaton only
when it is oscillating or when it dominates 
(and its intrinsic entropy thus vanishes),
because otherwise $\zeta_\sigma$ is not conserved.
In the early radiation dominated era,
the expression for the entropy perturbation is given in Eq.~(\ref{S}).
When $\sigma$ oscillates, $\rho_\sigma/\dot{\sigma}^2=1$ and $S$ is 
constant with the value
  $S = -2 \d \sigma_* /\sigma_*$.
Substituting in (\ref{transfert2}) one indeed recovers
(\ref{transfert_curvaton}).

\subsubsection{The secondary  inflaton limit}
\begin{figure}
\begin{center}
\includegraphics[width=4.8in]{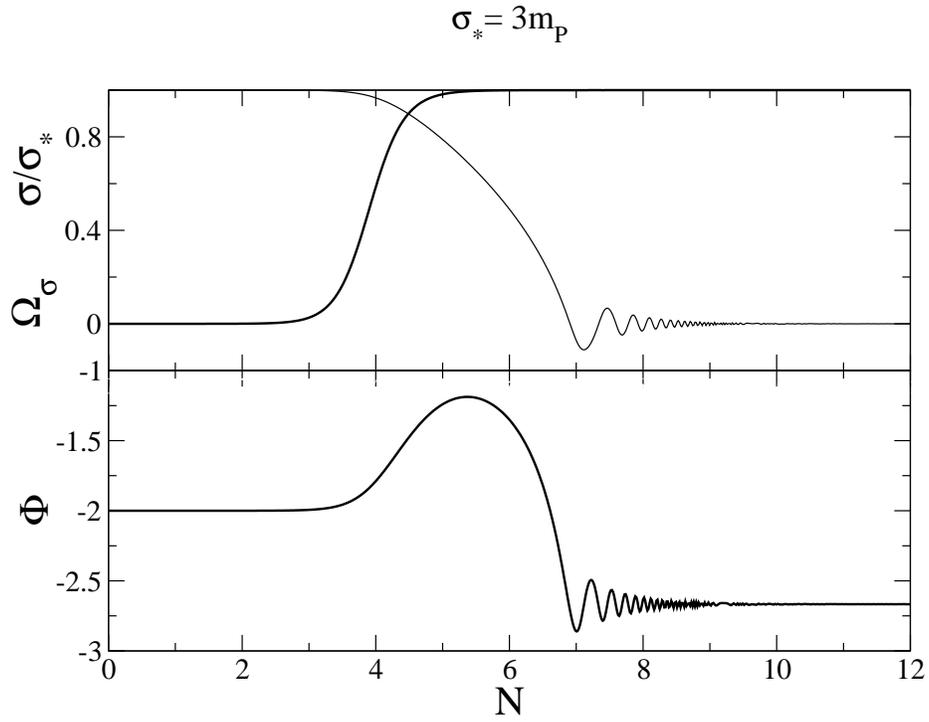}
\end{center}
\caption{The secondary  inflaton limit, for $\sigma_* = 3 \mP$.
Evolution of $\sigma$, its energy density fraction (upper part),
and of the curvature perturbation (lower part). The initial
conditions for the perturbations are $\delta \sigma_* = \mP$ and
$\Phi_* =-2$. }
\end{figure}

 Another limiting case corresponds to the situation
where $\sigma$ starts  oscillating only after it completely
dominates the universe. This implies that the curvaton $\sigma$ is
responsible for an additional phase of inflation: this is why we
call $\sigma$ the {\it secondary inflaton} in this situation.

We now attempt  to evaluate analytically the final curvature
perturbation. Let us consider again the  entropy perturbation $S$.
As mentioned already, $S$ is not constant because $\zeta_\sigma$
is not conserved. This can be checked explicitly with the expression 
\beq S =-2
\frac{\rho_\sigma}{\dot \sigma^2}
 \frac{\d \sigma_*}{\sigma_*} \, , \label{entropy_inflation}
\eeq 
already given in Eq.~(\ref{S}). 
Indeed, in the early
radiation dominated era, 
 using the explicit solution given in (\ref{sigma_RD1}), one
finds the behavior  $\rho_\sigma/\dot \sigma^2 \simeq
(25/8)(mt)^{-2}$.

In the subsequent phases dominated by the
curvaton, i.e.\ the secondary inflation and the oscillating phase,
the entropy $S$ remains constant. 
Although the expression (\ref{entropy_inflation}) is 
rigorously valid only during 
the phase when the curvaton is negligible, we can evaluate it at the 
transition  between the (first) radiation dominated era and
the curvaton dominated  era to get an estimate of the subsequently 
constant value of $S$.  

During the curvaton dominated phase,
$\sigma$
follows a slow-roll motion, characterized by \beq 3H^2\simeq {m^2
\sigma^2\over 2\mP^2}, \quad  \dot\sigma=-{m^2\sigma\over 3H},
\eeq and therefore, the coefficient $\rho_\sigma/\dot \sigma^2$
takes the constant value \beq \frac{\rho_\sigma}{\dot \sigma^2}
\simeq
 \frac{3}{4} \frac{\sigma_*^2}{\mP^2},  \label{factor}
\eeq where we have used the fact that $\sigma$ is essentially
frozen at its inflationary value $\sigma_*$ throughout the
radiation dominated era. 
Inserting this ratio in   Eq.~(\ref{entropy_inflation}), 
we get 
\beq 
S = -\frac{3 \sigma_*}{2
\mP^2} \d \sigma_*. 
\eeq 
Substituting the above value in
Eq.~(\ref{transfert2}), we then find for the   curvature perturbation
in the strongly oscillating phase 
\beq
\Phi_f={9\over 10}\Phi_* -{3\sigma_*\over 10 \mP^2}\d\sigma_*.
\eeq 
We have checked numerically that our analytical derivation
gives a very good approximation. This is illustrated in Fig.~3 for
two different values of $\Phi_*$ (and with $\d\sigma_*=\mP$).

\subsubsection{The general case}

\begin{figure}
\begin{center}
\includegraphics[width=4.8in]{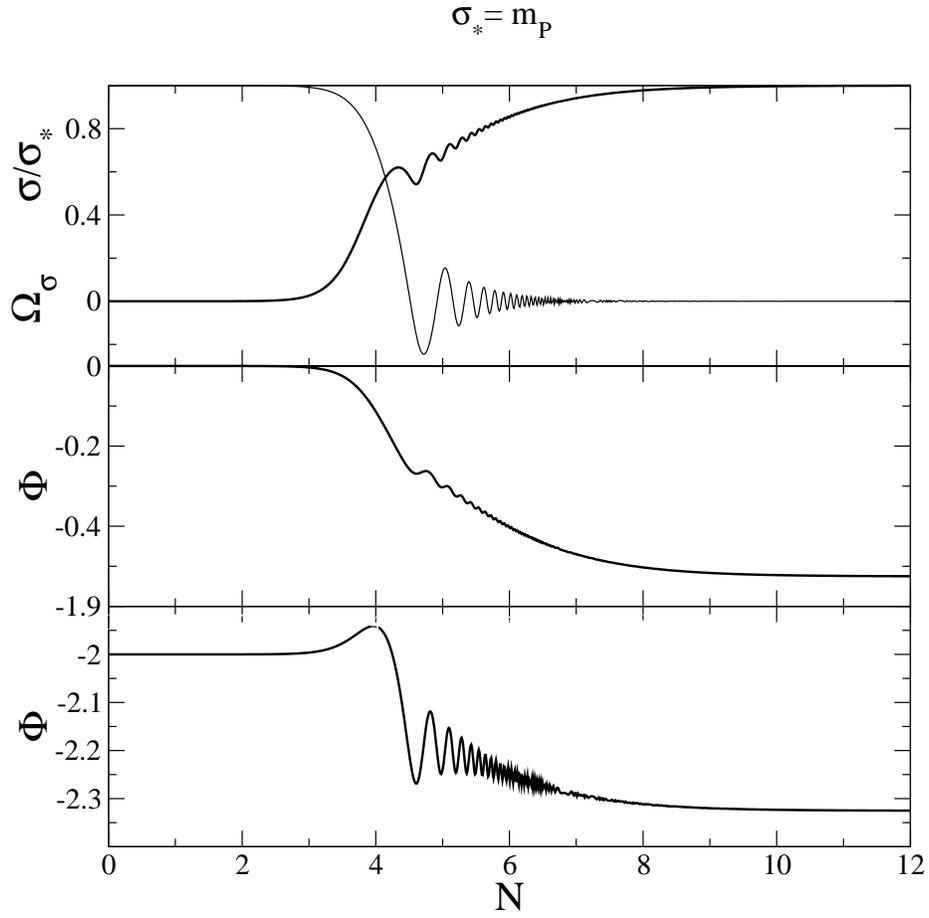}
\end{center}
\caption{The intermediate case, for $\sigma_* =  \mP$. Evolution
of $\sigma$, its energy density fraction (upper part), and of
the curvature perturbation (lower part) for two different initial
conditions: $\Phi_*=0$ and
for $\Phi_* =-2$ (both cases with $\delta \sigma_* = \mP$). }
\end{figure}

We now consider the general case, which includes the intermediate situations
 where the curvaton
starts oscillating at about the same time it becomes dominant.
We have  computed  the dependence of the
coefficient $\beta$ on the initial value $\sigma_*$, by solving
numerically the full system of equations  
for different values of   $\sigma_*$: the intermediate
case $\sigma_*=\mP$ is plotted in Fig.~3 as an illustration.

\begin{figure}
\begin{center}
\includegraphics[width=4.8in]{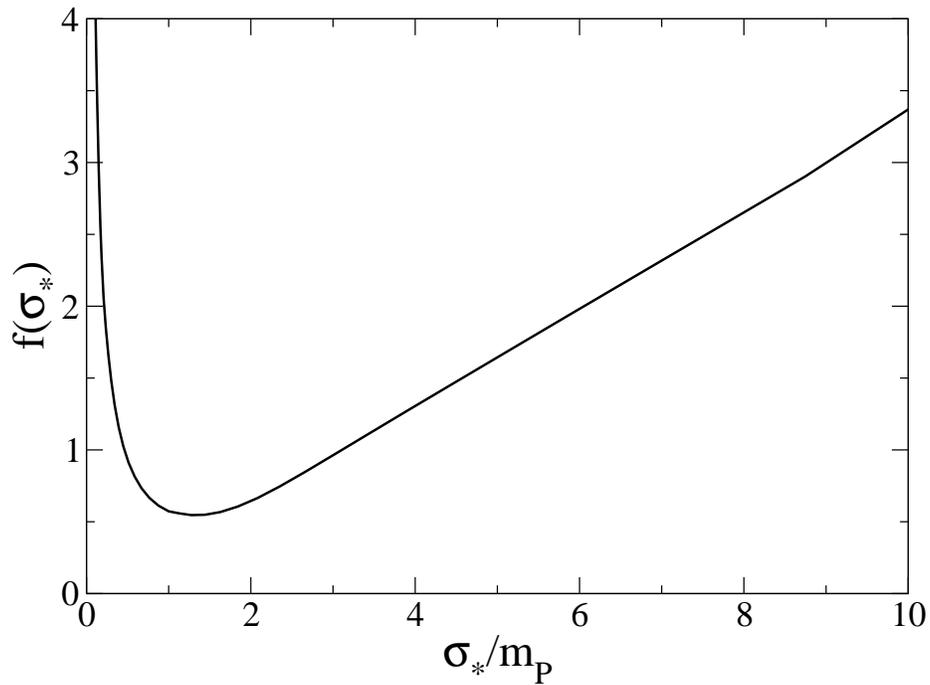}
\end{center}
\caption{The function $f(\sigma_*)$, which characterizes the
amplitude of the contribution of $\sigma$ to the curvature
perturbation. For $\sigma_* \ll
\mP$ one recognizes the $\propto 1/\sigma_*$  contribution of the
pure curvaton model. For $\sigma_* \gg \mP$, one recognizes the
 contribution of a secondary inflaton, proportional 
to $\sigma_*$
.}
\end{figure}

The resulting function, up to a normalization constant defined
just now, is plotted in Fig.~4. Instead of using  $\beta$, we
prefer to work with the coefficient that appears in the expression
for the curvature perturbation {\it after the decay of the
curvaton}, i.e.\ in the second radiation phase of our scenario,
identified with the radiation dominated era of the 
standard big bang model. The relation between the curvature
perturbation $\Phi_f$ at the end of the curvaton dominated phase
and the curvaton perturbation $\Phi_{RD}$ in the post-curvaton
radiation phase is  given by 
\beq 
\Phi_{RD}={10\over
9}\Phi_f, \eeq 
the transfer coefficient being simply due to the
transition from a matter dominated phase to a radiation
dominated phase. It is then convenient to define  the function $f$
 as
\beq 
\label{transfert_general} 
\Phi_{RD}=\Phi_* -{f(\sigma_*)\over \mP}
\delta \sigma_*\, . 
\eeq 
As we can see on Fig.~4,  for small
values of $\sigma_*$, we are in the standard curvaton regime with
\beq 
f(\sigma_*)= {4\over 9}{\mP\over \sigma_*},
\quad {\rm pure \ curvaton}, \label{pure_curvaton}
\eeq 
whereas for large values of
$\sigma_*$, we recover  the secondary inflaton limit
characterized by 
\beq 
f(\sigma_*)= {\sigma_*\over 3 \mP},
\quad{\rm secondary \ inflaton}. 
\eeq 
In the intermediate range,
$f$ reaches its minimum value, $f\simeq 0.55$ 
around $\sigma_*\simeq 1.2 \mP$.

\section{Observational consequences}

\subsection{Power spectrum of the primordial fluctuations}

So far, we have computed the primordial fluctuations, 
defined in the second radiation era  after the
decay of the curvaton, without specifying the initial curvature 
perturbation $\Phi_*$.
However, in the context of an inflationary era driven by 
a single slow-rolling scalar field, one can relate 
$\Phi_*$ to the fluctuation of the inflaton.  
The perturbations due to the inflaton can be computed following
the standard technique for a slow-rolling single field inflation
model (see e.g.\ \cite{Stewart}). One finds 
\beq
\Phi_*=-[1 -(1+3C) \epsilon + C \eta] 
{2\over 3\mP^2}\left. {V\over V'}\right|_* \d\phi_*, 
\eeq
where $V=V(\phi)$ is the potential of the inflaton,
$V'\equiv dV/d\phi$, $\epsilon$ and $\eta$ are the first two slow-roll
parameters,
\beq 
\epsilon={\mP^2 \over 2}\left({V'\over V}\right)^2,
\eeq
\beq
\eta=\mP^2{V''\over V}, 
\eeq 
and $C =-2+\ln 2 +\gamma \simeq -0.73$, $\gamma$ being
the Euler constant.
It is important to include the slow-roll corrections in the 
normalization of the curvature perturbation since these can be 
of the same order as the curvaton contribution.
The subscript $*$ 
means that all quantities have to be  evaluated at
Hubble crossing during inflation. 
Note that we are implicitly assuming that large scales 
that are observable today, i.e., those of order $H_0^{-1}$,
did not reenter the Hubble radius between the end of inflation
and the curvaton domination.

Substituting in our result of the
previous section, Eq.~(\ref{transfert_general}), the primordial curvature
fluctuations in the second radiation era are thus given by the
combination 
\beq 
\label{curvature}
\Phi_{RD}=- [1 -(1+3C) \epsilon + C \eta ]{2\over 3\mP^2}\left. {V\over
V'}\right|_* \d\phi_* -{1\over \mP}f(\sigma_*) \d\sigma_* . 
\eeq
The fluctuations $\d\phi_*$ and $\d\sigma_*$ must be seen as {\it
independent} random fields with the same amplitude for their power
spectra  and can be written as 
\beq 
\label{random}
\d\phi_*={H_*\over 2\pi}
e_\phi, \qquad \d\sigma_*={H_*\over 2\pi} e_\sigma, 
\eeq 
with
$\langle e_\phi e_\sigma \rangle = 0$. It is also worth
noticing
 that our result, in the secondary inflaton limit discussed before, with
$f(\sigma_*)= \sigma_*/(3 \mP)$ is in agreement with the
expression obtained in \cite{ps92} (see also \cite{l99})
 for the curvature perturbation
given in terms of the perturbations of the two scalar fields in a
double  inflation model.

Let us drop the subscript $*$.
Using (\ref{curvature}) and (\ref{random}), the amplitude of the 
 power spectrum for $\Phi$ is 
\def\f{{\tilde f}}
\beq
{\cal P}_\Phi= 
[1 + \f^2 \epsilon - 2 (1+3C) \epsilon + 2 C \eta] {H^2\over 18 \pi^2
\epsilon \mP^2} ,
\label{spectrum} 
\eeq
where we have introduced 
\beq
\f \equiv \frac{3}{\sqrt{2}} f . \label{deff}
\eeq
The contribution of the curvaton to the total curvature 
perturbation appears here on the same footing as the slow-roll corrections in
the amplitude of the inflaton power spectrum. However, when $\f \gg 1$
the curvaton contribution dominates over the slow-roll corrections and 
the term $-2(1+3C) \epsilon +2C\eta$ can be neglected.

The relative importance of the curvaton over the inflaton
perturbation is parametrized by 
\beq
R \equiv \frac{\f^2 \epsilon}{1 - 2 (1+3C) \epsilon + 2 C\eta}  
\simeq \f^2 \epsilon.
\eeq  
This is given by the product of two  parameters that depend on 
two a priori independent physical processes: 
$\epsilon$, which must be smaller than one, is determined by 
the inflationary scenario, while   
$\tilde f^2 \gsim 1.36$ is determined 
by the curvaton expectation value.
 The slow-roll parameter $\epsilon$ is typically of order 
$\sim 1/60$ in chaotic models of inflation but it can be much smaller 
in other  inflationary models. 
Therefore, there are cases where the curvaton 
value during inflation is small, $\sigma \ll \mP$, 
and yet the perturbations generated 
by the inflaton are of the order of or larger than 
the curvaton perturbations, i.e.\ $R \lsim 1$.

When
the curvaton perturbation is absent, i.e.\ for $\f=0$, we
recover  the pure inflationary spectrum  (see
e.g.\ \cite{Stewart}), 
\beq 
{\cal P}_\Phi= [1 - 2( 1+3C) \epsilon + 2 C \eta] 
{H^2\over 18 \pi^2
\epsilon \mP^2} . 
\eeq 
Conversely, when the curvaton
perturbation dominates the power spectrum, i.e.\ for 
$R \gg 1$ and $\sigma \ll  \mP$, we recognize the spectrum
of perturbations of the curvaton \cite{lw02}, \beq {\cal P}_{\Phi}
= \frac{4 H^2}{81 \pi^2
 \sigma^2},
\eeq
where we have used $\f=(2\sqrt{2}/3)(\mP/\sigma)$ 
from Eqs.~(\ref{pure_curvaton}) 
and (\ref{deff}).

The presence of mixed  curvaton and inflaton perturbations also
affects the scalar spectral  index, which is defined as 
\beq 
n_s-1
\equiv  \frac{d \ln {\cal P}_{\Phi}}{d \ln k}. 
\eeq 
Inserting in
this expression the spectrum (\ref{spectrum}) and assuming that
the curvaton mass is much smaller than the inflaton mass, $m \ll
V''$, one finds 
\beq 
n_s-1 = -2 \epsilon + \frac{2 \eta - 4
\epsilon } {1+ \f^2 \epsilon}+
[-2(7+12C) \epsilon^2 
+2(3+8C) \epsilon \eta -2C \xi^2], \label{spectral_index} 
\eeq
where we have introduced the second-order slow-roll parameter
\beq 
\xi^2 \equiv \mP^4 \frac{V'V'''}{V^2} . 
\eeq 
In the square bracket we have isolated terms which are second order in the 
slow-roll parameters. These can be as important as the curvaton contribution
if $\f \sim 1$.

Neglecting second order terms, in the case of pure inflation,
$\f^2 =0$, one recovers the well-known result 
\beq
n_s-1=2\eta-6\epsilon, 
\eeq 
whereas in the case of a pure
curvaton, $\f^2 \epsilon \gg1$, 
one gets \cite{lw02} 
\beq
n_s-1=-2\epsilon. 
\eeq

The power spectrum for gravitational  waves is given by the
usual expression (see e.g.\ \cite{Stewart}) 
\beq {\cal P}_T =
[1 -2(1+C) \epsilon] \frac{2 H^2}{\pi^2 \mP^2} . 
\eeq
 Using
(\ref{spectrum}), we thus find for the tensor-scalar ratio
 \beq
r \equiv \frac{4}{9} \frac{{\cal P}_T}{{\cal P}_{\Phi}}=
 \frac{16 \epsilon}{1+\f^2 \epsilon}  [1 +4C \epsilon - 2C\eta] 
\label{ratio}
\eeq 
(we have defined $r$ as in  \cite{peiris}). Neglecting second 
order terms in the slow-roll parameters, for $\f=0$ we
recover $r=16 \epsilon$, the tensor-scalar ratio for
a pure inflationary model. For a pure curvaton perturbation,
$\f^2\epsilon \gg 1$, one finds $r\simeq 16/\f^2\simeq 18(\sigma_*/\mP)^2$. 
In the usual curvaton context, where $\sigma_* \ll \mP$, this implies
 that gravity waves are negligible
\cite{luw}.

Adding a curvaton perturbation
to a pure inflationary perturbation always decreases the tensor-scalar
ratio $r$.  When $\f$ is large, 
the effect on the scalar index depends on the sign of
$2 \eta - 4 \epsilon$: if it is negative  (this is the case
for the so-called  small field and chaotic models \cite{kinney}),
the scalar index is increased, which means that the spectrum is
bluer; if it is positive, the effect is opposite.

Models of inflation are often classified according to the value of their 
first two slow-roll parameters $\epsilon$ and $\eta$ \cite{kinney}.
Since for $\f =0$ there is a one-to-one 
correspondence between the $(\epsilon,\eta)$-plane and the $(n_s,r)$-plane,
it is customary to use the latter in order to easily confront models 
with observations. In the mixed case, $\f \neq 0$, 
the presence of the curvaton
introduces a degeneracy between these two planes, which depends 
on the parameter $\sigma_*$.
In the next
subsection we will use Eqs.~(\ref{spectral_index}) and
(\ref{ratio}) to show the effects of this degeneracy for the special case of
$\lambda \phi^4$-inflation.

Let us  now discuss  the consistency relation. We restrict 
the discussion to the large $\f$ case and we neglect terms which are 
second order in the 
slow-roll parameters. The tensor spectral
index being given by the usual expression 
\beq 
n_T \equiv {d \ln
{\cal P}_T\over d \ln k} = - 2 \epsilon \, ,
\eeq
 the consistency relation between the
tensor-scalar ratio and the tensor spectral index is modified into 
 \beq r =
\frac{-8  n_T}{1- \f^2 n_T/2}. \label{consistency} \eeq 
For large
$\f$ we find a modification of the standard consistency relation
valid for a  pure single-field inflation. This represents a distinctive
feature of mixed curvaton and inflaton perturbations, which 
breaks the degeneracy of
mixed  models.
Indeed, this modification
depends solely on the parameter $\sigma_*$, which determines the
value of $\f$: since the relation (\ref{consistency}) 
does not depend on a particular
inflaton potential, it can be seen as a way of measuring
$\sigma_*$.

For completeness,  we briefly  discuss the running of the scalar
spectral index. We find
 \beq \frac{d n_s}{d \ln k}
= 4 \epsilon (\eta - 2 \epsilon) - \frac{16\epsilon^2 -12
\epsilon \eta + 2 \xi^2}{1 + \f^2 \epsilon} + \frac{4 \f^2
\epsilon (\eta - 2 \epsilon)^2}{(1 + \f^2 \epsilon)^2}.
\eeq
For $\f =0$ we
recover the pure inflationary running \cite{peiris}, \beq \frac{d
n_s}{d \ln k} = -24 \epsilon^2 +16 \epsilon \eta -2 \xi^2 ,
\eeq 
while for
$\f^2 \epsilon \gg 1$ we find the running for a pure curvaton
spectral index, \beq \frac{d n_s}{d \ln k} = 4 \epsilon (\eta
- 2 \epsilon). \eeq

In the next subsection, we will illustrate the general formulas derived here 
by considering the particular case of an inflaton with a quartic potential. 

\subsection{Mixed 
perturbations with  $V(\phi)=\lambda \phi^4$}

The quartic potential $V = \lambda \phi^4$ is a simple
inflationary potential which has attracted some attention
lately as it lies in a region excluded by the WMAP data analysis
\cite{peiris,Barger,Leach}. For this  particular inflationary
potential,  we now illustrate how mixed curvaton and inflaton
perturbations can 
be constrained by the data and how the
observational predictions of a pure inflationary model change  in
the presence of a curvaton.

The slow-roll parameters for this potential are 
 \beq
\epsilon = 8 \frac{\mP^2}{\phi^2}, \quad \eta = 12
\frac{\mP^2}{\phi^2}, \quad \xi^2 = 96 \frac{\mP^4}{\phi^4}. 
\label{slowroll_phi4}
\eeq
The number of $e$-foldings before the end of inflation when 
the perturbations at 
our present Hubble scale exited the Hubble 
radius   during inflation is given by 
\beq
N_* = \ln {a_{\rm end}
\over  a_*} = {\phi_*^2 \over 8\mP^2}, \label{efoldings_phi4}
\eeq
 where we have assumed that $\phi_{\rm
end} \ll \phi_*$. Using this expression we can write the slow-roll
parameters in terms of the number of $e$-foldings $N_*$.

Determining the appropriate number of $e$-foldings $N_*$ is very
important, since it tells us  which part of the potential corresponds 
to the present observable fluctuations. 
This depends on the history of the universe from the
end of inflation until today. In particular, it is very sensitive
on the duration of  the reheating phase: this is usually the
largest uncertainty in its estimation. However, in the case of
pure  inflation with a quartic potential, $N_*$ can be evaluated
very accurately because the expansion during reheating is the same as in
a radiation dominated era. Indeed, the energy density of an
oscillating scalar field in a quartic potential behaves as
radiation, and the duration of the reheating phase is thus no longer
important. In this case, the number of $e$-foldings can be 
evaluated precisely  \cite{Liddle} and is 
 \beq
N_*^{\rm quartic} \simeq 64. \label{efolds} 
\eeq

\def\CD{{\rm CD}}
\def\dec{{\rm dec}}
\def\osc{{\rm osc}}

This  number  assumes radiation domination
from the end of inflation until equality. It does not take into
account a possible intermediate phase of curvaton domination
between reheating and nucleosynthesis, which lasts 
\beq
\Delta N=\frac{1}{3} \ln \frac{\rho_\CD}{\rho_\dec} 
\eeq
$e$-foldings, where $\rho_\CD$ is the total energy density 
 at curvaton and radiation equality and 
$\rho_\dec \simeq 3 \Gamma^2 \mP^2$ is the energy density of the universe
when the  curvaton decays with decay rate $\Gamma$. Curvaton domination
thus {\it decreases} the number of $e$-foldings 
which becomes 
 \beq
N_* \simeq
 64  - \frac{1}{12}
\ln \frac{\rho_\CD}{\rho_\dec}.
\label{correction}\eeq

Let us compute this correction in more details. If $\sigma_* \sim
\mP$ the domination phase starts with a small period of inflation.
When the curvaton and  radiation energy densities are the same, 
we have for the total energy density
\beq \rho_\CD =2\rho_\sigma\simeq
 m^2 \sigma_*^2 \label{inflation_domination}. 
\eeq 
In
this case we obtain 
\beq 
N_* \simeq  64.1  - \frac{1}{6} \ln
\frac{m}{\Gamma} -\frac{1}{6}
 \ln \frac{\sigma_*}{\mP}, \quad {\rm for} \ \ \sigma_* \sim \mP.
 \eeq

It is also interesting to consider  the pure curvaton case corresponding to
$\sigma_* \ll \mP$.  In this case Eq.~(\ref{inflation_domination}) 
does not apply, but we can still compute the ratio 
${\rho_\dec/{\rho_\CD}}$. Once the curvaton starts
oscillating at $H \sim m$,  we have 
\beq 
\rho_{\sigma} \simeq m^2 \sigma_*^2
\left( \frac{a_\osc}{a} \right)^3, 
\quad {\rho_{r}} \simeq 3 m^2
\mP^2 \left( \frac{a_\osc}{a} \right)^4  , 
\eeq 
where $a_\osc$ is the scale factor at the onset of the curvaton oscillating
period. This implies 
\beq 
a_\CD/a_\osc \simeq 3 \mP^2 /\sigma_*^2.
\label{at_equality} 
\eeq 
On the other hand, when the  curvaton decays ($H
\sim \Gamma$), it dominates the universe and  
\beq m^2
\sigma_*^2 \left( \frac{a_\osc}{a_\dec} \right)^3
\simeq 3 \Gamma^2 \mP^2. \label{at_decay} 
\eeq 
On combining
Eqs.~(\ref{at_equality}) and (\ref{at_decay}) we can compute the
correction factor in Eq.~(\ref{correction}), 
and we obtain finally 
\beq 
N_* \simeq 64.4
 - \frac{1}{6} \ln \frac{m}{\Gamma} - \frac{2}{3} \ln
\frac{\sigma_*}{\mP}, \quad {\rm for} \ \ \sigma_* \ll \mP .
\eeq

The ratio $m/\Gamma$ depends on the specific  model for the curvaton. 
It parameterizes the
duration of the curvaton oscillating phase  and represents here the main
uncertainty, although limited by the  prefactor $1/6$. 
Also the last term   can be
important for very small $\sigma_*$: the smaller $\sigma_*/\mP$ is, the
later the curvaton starts dominating the universe, thus shortening the
duration of the curvaton domination phase. This  increases
the number of $e$-foldings and  partially compensates  the $\ln
(m/\Gamma)$ correction.

\begin{figure}
\begin{center}
\includegraphics[width=4in]{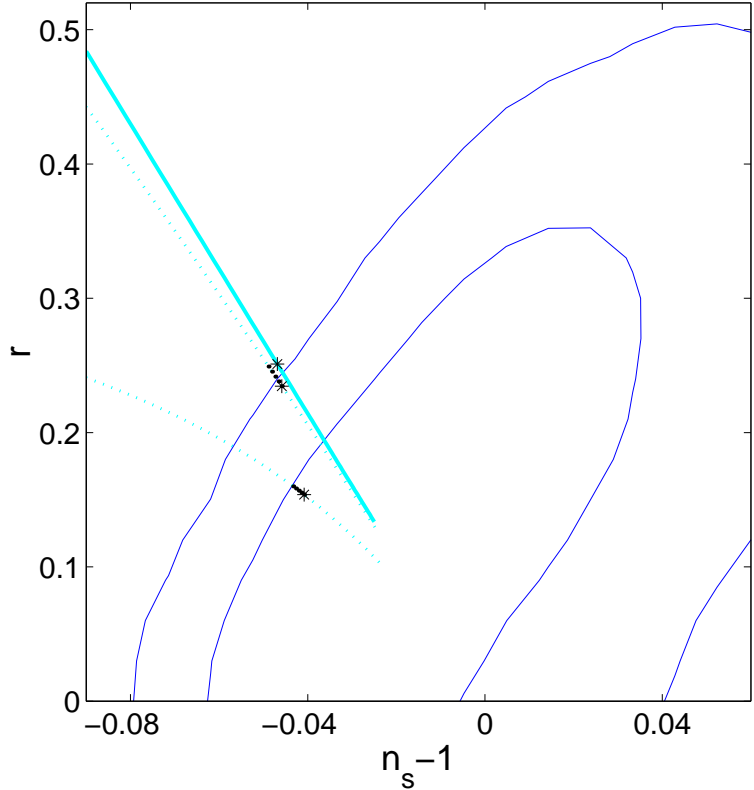}
\end{center}
\caption{Two-dimensional likelihood contours at $68 $\% and $95
$\% confidence level of the WMAP data on the $(n_s,r)$-plane, as
compared to the predictions of a pure inflationary model (solid
$N$-trajectory), a mixed model with 
$f=1$ 
(upper dotted $N$-trajectory), corresponding to $\sigma_* \sim 0.5 \mP$, and one with 
$f=3$ (lower dotted $N$-trajectory), 
corresponding to $\sigma_* \sim 0.1 \mP$, respectively.
The stars denote $N=64$ on the $N$-trajectories.  The interval $60
\le N \le 64$ is represented by bold dots on the $N$-trajectories
of the mixed models. The likelihood contours are from the analysis
of S. Leach and A. Liddle \cite{Leach}.} \label{WMAP}
\end{figure}

Now we have all the ingredients to study the  predictions of
mixed models as compared to a pure quartic inflationary model. 
We use Eqs.~(\ref{slowroll_phi4}) and (\ref{efoldings_phi4}) 
to express the slow-roll parameters $\epsilon$ and $\eta$
in terms of the number of $e$-foldings $N_*$. 
Furthermore, on using Eqs.~(\ref{spectral_index}) 
and (\ref{ratio}),
we can express the observables $n_s$ and $r$ in terms of $N_*$,
and we can compare the predictions on the $(n_s,r)$-plane of a  
pure quartic model and mixed models with cosmological data.

In Fig.~\ref{WMAP} the two dimensional likelihood contours at $68 $\%
and $95 $\%  confidence level of the WMAP data (combined with 
other data as from the analysis of \cite{Leach})
are shown on the
$(n_s,r)$-plane. 
For a pure inflationary model, the
$N_*=64$ realization is marginally excluded by the data. When
adding the perturbation of the curvaton, the predictions on the
$(n_s,r)$-plane are changed and the $N_*=64$ realization of the
inflationary model can be safely included into the $95 $\%
confidence level contour for a mixed model with $\sigma_* \lsim
0.5 \mP$, and into the $68 $\% confidence level contour for a
mixed model with $\sigma_* \lsim 0.1 \mP$. However, 
one must also take into account the fact that the number of 
$e$-foldings $N_*$  decreases because of  the phase of curvaton
domination, according to (\ref{correction}).
 As an illustration of this effect, we consider the
interval $ 60 \le N_* \le 64$ on the curvaton trajectory and we
plot it in Fig.~\ref{WMAP}. Decreasing the number of $e$-foldings
tends to exclude the model.

As anticipated in the introduction, the extraction of information
from the data is limited now by the introduction of a new field,
the curvaton, and  a corresponding new parameter, the curvaton
expectation value during inflation. The degeneracy of the data can
be broken by making use of the consistency relation,
Eq.~(\ref{consistency}), although 
we are still far from measuring
$n_T$ \cite{Knox}.

\section{Conclusion}
In the present work, we have explored one particular facet of the 
curvaton mechanism: we have considered the case where 
primordial 
curvature perturbations generated by the curvaton are of the same order 
of magnitude as the traditional inflaton-generated curvature perturbations. 
This can happen when 
the vacuum expectation 
value for the curvaton is of the order of the Planck mass during inflation. 
This situation can  occur in some scenarios, such as in the string axion 
scenario discussed in \cite{dlnr03}.

Independently of the specific model of inflation, we have obtained the 
expression for the ``primordial'' curvature perturbation, given as a 
linear combination of the post-reheating curvature perturbation (i.e.\ 
 generated during inflation) and of the curvaton perturbation. 
In terms of the curvature perturbation on the uniform energy density
hypersurface this is given as
\beq
\zeta= \left[ { 1 \over \sqrt{ \epsilon}} 
  e_\phi + {\tilde f}(\sigma)   e_\sigma \right] 
\frac{H}{2 \sqrt{2} \pi \mP}, 
\quad \langle e_\phi e_\sigma \rangle = 0,
\eeq
where $H$, $\sigma$, and $\epsilon$ are the Hubble parameter, 
the vacuum expectation value of the curvaton, and the first slow-roll 
parameter, all evaluated at Hubble crossing during inflation, respectively
[see Eq.~(\ref{curvature})].
This expression is only valid for large $\f$, 
otherwise corrections which
are second order in the slow-roll parameters must be introduced.
The function $\tilde f(\sigma)$ depends on the initial background 
value for the curvaton and interpolates between two regimes, which have 
been investigated previously in the literature: 
the standard curvaton 
regime, where the curvaton starts oscillating long before it dominates, and
for which $\tilde f = (2\sqrt{2}/3)(\mP/\sigma)$, and 
the secondary  inflaton regime, where the ``curvaton'' is still frozen when 
it dominates and thus produces an additional period of inflation, 
and $\tilde f = (1/\sqrt{2})(\sigma/\mP)$.

In the case of slow-roll inflation, we have computed the spectral 
amplitude and index associated with  curvature perturbations of mixed 
origins and shown how the resulting expressions nicely interpolate 
between the pure inflaton and pure curvaton cases. As an illustration, 
we have considered the case of a quartic potential for the inflaton
and shown how allowing for a mixing with a curvaton changes
the usual predictions. 

The results derived here  are quite generic and can be applied to any 
inflationary model. Although  we have considered, 
for simplicity,  only a quadratic potential for the curvaton, 
which is a good approximation near a local minimum of most potentials, it is
in principle straightforward to extend our results to other potentials. 
 As we have shown, a generic consequence of mixed perturbations is 
a modification of the usual consistency relation, which gives some hope to 
be able to distinguish  observationally between the standard inflationary
scenario and its mixed version.

\vskip 1cm
{\bf Acknowledgements}
We wish to thank David Lyth, J\'er\^ome Martin and David Wands for 
stimulating discussions and instructive comments.  F.V. acknowledges 
financial support from the Swiss National Science Foundation.

\end{document}